\documentclass[12pt,reqno]{amsart}
\usepackage{amsmath}
\usepackage{amsfonts}
\usepackage{amsthm}
\usepackage{amssymb}
\usepackage{hyperref}

\theoremstyle{definition}

\numberwithin{equation}{section}
\begin{document}

\title{Pricing barrier options with discrete dividends}
\author{D. Jason Gibson}
\author{Aaron Wingo}

\address{Department of Mathematics and Statistics,
Eastern Kentucky University, KY 40475, USA}
\email{jason.gibson@eku.edu}

\address{Department of Mathematics and Statistics,
  Eastern Kentucky University, KY 40475, USA}
\email{aaron\_wingo@mymail.eku.edu}



\thanks{Both authors thank Tian-Shyr Dai and Chun-Yuan Chiu
for kindly granting permission to use their barrier option pricing data.}

\date{\today}

\begin{abstract}
  The presence of discrete dividends complicates the derivation
  and form of pricing formulas even for vanilla options.
  Existing analytic, numerical, and theoretical approximations
  provide results of varying quality and performance.
  Here, we compare the analytic approach, developed and effective
  for European puts and calls, of Buryak and Guo
  with the formulas, designed in the context
  of barrier option pricing, of Dai and Chiu.
\end{abstract}

\maketitle

\section{Introduction}

Following Buryak and Guo \cite{bg}, we focus on the analysis
of a stock process $S_t$ that jumps down by dividend amounts $d_i$
at times $t_i$. At non-dividend times, $S_t$ follows a geometric
Brownian motion with flat volatility $\sigma$. In this context, we have
\begin{equation}
  dS_t=\left(rS_t-\sum_{0<t_i\le T}d_i\, \delta(t-t_i)\right)\, dt+\sigma
  S_t\, dW_t,\end{equation}
where $r$ is the risk-free interest rate, $\delta$ is the Dirac
delta function, and $W_t$ is a Wiener process. (The book by Hull
\cite{hull2} serves as a standard reference on these matters.)
The Black-Scholes partial differential equation
\begin{equation}
  \frac{\partial V}{\partial t}-rV+rS\frac{\partial V}{\partial S}
  +\frac{1}{2}\sigma^2S^2\frac{\partial^2V}{\partial S^2}=0
\end{equation}
models the dynamics of the option price. Here, $S$ denotes the (spot)
asset price, $\sigma$ denotes volatility, and $r$ denotes
the flat interest rate.

The presence of discrete dividends complicates the derivation
and form of pricing formulas even for vanilla options, let alone
barrier options or more exotic instruments.
Existing analytic, numerical, and theoretical approximations
provide results of varying quality and performance.
One possibility, discussed by Frishling \cite{frishling}
and used under the name \lq\lq Model 1\rq\rq\
for the sake of comparison by Dai and Chiu \cite{dai},
holds that the difference between the stock price and
the present value of future dividends over the life of the option
follows a lognormal diffusion process. More involved approaches,
like those of Buryak and Guo \cite{bg} and Dai and Chiu \cite{dai},
respectively, allow more sophisticated and sensitive
incorporation of factors influencing the option price
(volatility, barriers, etc.). Numerical approximations,
including Monte-Carlo methods, lattice methods, and
Crank-Nicolson schemes, often provide benchmarks
for other methods, whether sophisticated or naive.

In \S 2, we describe the analytic approximations
from Buryak and Guo \cite{bg}: the spot volatility adjusted,
strike volatility adjusted, hybrid, and hybrid volatility adjusted
approximations. The latter approximation originates in
the paper \cite{bg}, where it performed well in pricing
calls and puts. In \S 3, we set up the analytic pricing formula,
valid in the absence of
discrete dividends,
for up and out barrier options. In \S 4, the performance
of the Buryak and Guo hybrid volatility adjusted approximation,
adapted to the setting of barrier options, can be seen
in charts that incorporate data from Dai and Chiu \cite{dai}.
In \S 5, we briefly sketch directions for further work
on these and related problems.

\section{The analytic approximations}

The conventional Black-Scholes formulas
\begin{align}
  \begin{split}\label{call_put}
  C&=S_0\,\Phi(b_1)-K\exp(-rT)\Phi(b_2),\\
  P&=K\exp(-rT)\Phi(b_2)-S_0\,\Phi(-b_1)
  \end{split}
\end{align}
do not provide for the possibility of stocks
with discrete dividends. Here, $C$ and $P$
denote Call and Put, respectively. We also
have stock (spot) price $S_0$, strike price $K$,
time $T$ to maturity for the option, $\Phi$ the
cumulative Gaussian distribution function, and
$b_i$ that satisfy
\begin{align}
  \begin{split}\label{bs_parameters}
  b_1&=\frac{1}{\sigma\sqrt{T}}\left(\ln\frac{S_0}{K}+
  \left(r+\frac{\sigma^2}{2}\right)T\right)\\
  b_2&=b_1-\sigma\sqrt{T}.
  \end{split}
\end{align}

\subsection{Spot volatility adjusted approximation}

Beneder and Vorst \cite{bv} use an approximation
that, roughly speaking, adjusts some of the Black-Scholes
parameters and then adjusts the volatility to
refine the correction. To incorporate the dividend information,
one might subtract the present value of the dividends
\begin{equation}
  D = \sum_{0<t_i\le T} d_i\exp(-rt_i)
\end{equation}
from $S_0$, producing the adjusted value
\begin{align}
  \begin{split}\label{spot_adjustment}
    \tilde S_0&=S_0-D\\
    &=S_0-\sum_{0<t_i\le T} d_i\exp(-rt_i).
  \end{split}
\end{align}

They observed that, provided the local volatility (of a stock
process with discrete dividends) is constant, the
process without dividend-induced jumps should then
have non-constant local volatilities
\begin{equation}
\tilde \sigma_S(S,D,t)=\sigma(T)\frac{S}{S-D_j^{(S)}},
\end{equation}
where
\begin{equation}
  D_j^{(S)}=\sum_{i=j(t)}^N d_i\exp(-rt_i),
\end{equation}
with $N$ being the number of dividend payments in $(0,T)$
and the sum restricted to include only those payments
occurring after time $t$, with $j(t)$ the index of the first
dividend payment at or after time $t$.

In light of this volatility adjustment, the corresponding
variance can be averaged on $(0,T)$, yielding
\begin{equation}
  \overline{\sigma}_S=
  \sigma\sqrt{
    \left(\frac{S}{S-D_1^{(S)}}\right)^2\frac{t_1}{T}
    +
    \sum_{1<j\le N}\left(\frac{S}{S-D_j^{(S)}}\right)^2\frac{t_j-t_{j-1}}{T}
    +
    \frac{T-T_N}{T}
    },
\end{equation}
with $t_N$ being the time of the last dividend payment in $(0,T)$.

Replacing $S_0$ by $\tilde S_0$ and $\sigma$ by
$\overline{\sigma}_S$ in \eqref{call_put}
and in \eqref{bs_parameters} yields
\begin{align}
  \begin{split}\label{call_put_div}
  C&=\tilde S_0\,\Phi(b_1)-K\exp(-rT)\Phi(b_2),\\
  P&=K\exp(-rT)\Phi(b_2)-\tilde S_0\,\Phi(-b_1)
  \end{split}
\end{align}
and
\begin{align}
  \begin{split}\label{bs_parameters_div}
  b_1&=\frac{1}{\overline{\sigma}_S\sqrt{T}}\left(\ln\frac{\tilde S_0}{K}+
  \left(r+\frac{\overline{\sigma}_S^2}{2}\right)T\right)\\
  b_2&=b_1-\overline{\sigma}_S\sqrt{T}.
  \end{split}
\end{align}
Call the above scheme the spot volatility adjusted approximation,
the spot VA approximation.

\subsection{Strike volatility adjusted approximation}
Buryak and Guo \cite{bg} introduce a different approximation
based on the set-up of Beneder and Vorst \cite{bv}.
First, following Frishling's description \cite{frishling}
of a strike approximation, they
modify the strike price from $K$ to $\tilde K$
by setting
\begin{equation}
      \tilde{K}=K+\sum_{0<t_i\le T} d_i\exp(r(T-t_i)),
\end{equation}
a natural analog of \eqref{spot_adjustment}.

Next, the volatilities get adjusted by considering
the non-constant local volatilities
\begin{equation}
\tilde \sigma_K(S,D,t)=\sigma(T)\frac{S}{S+D_j^{(K)}},
\end{equation}
where
\begin{equation}
  D_j^{(K)}=\sum_{i=1}^{j(t)} d_i\exp(-rt_i),
\end{equation}
with $N$ being the number of dividend payments in $(0,T)$
and the sum restricted to include only those payments
occurring before time $t$, with $j(t)$ the index of the first
dividend payment at or before time $t$.
With this volatility adjustment, the corresponding
variance can be averaged on $(0,T)$, yielding
\begin{equation}
  \overline{\sigma}_K=
  \sigma\sqrt{
    \frac{t_1}{T}
    +
    \sum_{1\le j<N}\left(\frac{S}{S+D_j^{(K)}}\right)^2\frac{t_{j+1}-t_j}{T}
    +
    \left(\frac{S}{S+D_N^{(K)}}\right)^2\frac{T-t_N}{T}
    },
\end{equation}
with $t_N$ being the time of the last dividend payment in $(0,T)$.

Replacing $K$ by $\tilde K$ and $\sigma$ by
$\overline{\sigma}_K$ in \eqref{call_put}
and in \eqref{bs_parameters} yields
\begin{align}
  \begin{split}\label{call_put_div_K}
  C&=S_0\,\Phi(b_1)-\tilde{K}\exp(-rT)\Phi(b_2),\\
  P&=\tilde{K}\exp(-rT)\Phi(b_2)-S_0\,\Phi(-b_1)
  \end{split}
\end{align}
and
\begin{align}
  \begin{split}\label{bs_parameters_div_K}
  b_1&=\frac{1}{\overline{\sigma}_K\sqrt{T}}\left(\ln\frac{S_0}{\tilde{K}}+
  \left(r+\frac{\overline{\sigma}_K^2}{2}\right)T\right)\\
  b_2&=b_1-\overline{\sigma}_K\sqrt{T}.
  \end{split}
\end{align}
Call the above scheme the strike volatility adjusted approximation,
the strike VA approximation.

\subsection{Hybrid approximation}
Bos and Vandermark \cite{bos} offered a different approximation,
one supported with some theoretical analysis. Specifically,
take
\begin{align}
  \begin{split}\label{hybrid_bsm}
  C&=\overline S_0\,\Phi(b_1)-\overline K\exp(-rT)\Phi(b_2),\\
  P&=\overline K\exp(-rT)\Phi(b_2)-\overline S_0\,\Phi(-b_1),
  \end{split}
\end{align}
with
\begin{align}
  \begin{split}
    \overline S_0 &= S_0-D_S\\
    \overline K   &= K + D_K \exp(rT).
  \end{split}
\end{align}
Here,
\begin{align}
  \begin{split}
    D_S &= \sum_{0<t_i\le T} \frac{T-t_i}{T}\, d_i \exp(-rt_i)\\
    D_K &= \sum_{0<t_i\le T} \frac{t_i}{T}\, d_i\exp(-rt_i).
  \end{split}
\end{align}
In contrast with the Spot VA and Strike VA approximations
described above, this method does not adjust the volatility.
Call the above scheme the hybrid approximation.

\subsection{Hybrid volatility adjusted approximation}\label{hybrid}
A new method described by Buryak and Guo takes
the Hybrid approximation above as a starting point,
but then also adjusts the volatilities in a manner
related to the volatility adjustment schemes
mentioned earlier. A key difference between
this new method and those other methods lies
in the individual treatment of the $D_S$ and $D_K$
terms, where the discounted dividend stream $D$
satisfies $D=D_S+D_K$.

Specifically, the volatilities get adjusted by considering
the non-constant local volatilities
\begin{equation}
\tilde \sigma_S(S,D,t)=\sigma(T)\frac{S}{S-D_j^{(S)}},
\end{equation}
where
\begin{equation}
  D_j^{(S)}=\sum_{i=j(t)}^N \frac{T-t_i}{T}d_i\exp(-rt_i),
\end{equation}
and
\begin{equation}
\tilde \sigma_K(S,D,t)=\sigma(T)\frac{S}{S+D_j^{(K)}},
\end{equation}
where
\begin{equation}
  D_j^{(K)}=\sum_{i=1}^{j(t)} \frac{t_i}{T}d_i\exp(-rt_i).
\end{equation}
Here, with $N$ being the number of dividend payments in $(0,T)$,
the spot sum gets restricted to include only those payments
occurring after time $t$, with $j(t)$ the index of the first
dividend payment at or after time $t$, and
the strike sum restricted to include only those payments
occurring before time $t$, with $j(t)$ the index of the first
dividend payment at or before time $t$.

In both instances, the corresponding
variance can be averaged on $(0,T)$, 
\begin{align}
  \begin{split}
  \overline{\sigma}_S &=
  \sigma\sqrt{
    \left(\frac{S}{S-D_1^{(S)}}\right)^2\frac{t_1}{T}
    +
    \sum_{1<j\le N}\left(\frac{S}{S-D_j^{(S)}}\right)^2\frac{t_j-t_{j-1}}{T}
    +
    \frac{T-T_N}{T}
  }\\
  &= \sigma(1+\varepsilon_S^{(h)}),
  \end{split}
\end{align}
and
\begin{align}
  \begin{split}
  \overline{\sigma}_K &=
  \sigma\sqrt{
    \frac{t_1}{T}
    +
    \sum_{1\le j<N}\left(\frac{S}{S+D_j^{(K)}}\right)^2\frac{t_{j+1}-t_j}{T}
    +
    \left(\frac{S}{S+D_N^{(K)}}\right)^2\frac{T-t_N}{T}
  }\\
  &= \sigma(1-\varepsilon_K^{(h)}),
  \end{split}
\end{align}
with $t_N$ being the time of the last dividend payment in $(0,T)$.

Finally, set
\begin{equation}
  \overline \sigma_H = \sigma(1+\epsilon_S^{(h)})(1-\epsilon_K^{(h)}),
\end{equation}
and use the set-up described by \eqref{hybrid_bsm}.
For calls, this approximation then requires no further
adjustments. Puts require further adjustment.
(See the Buryak-Guo \cite{bg} discussion of Liquidator
and Survivor dividend policies, based on considerations
described in Haug \cite{haug}.) Call this scheme the
hybrid VA approximation.

\section{Barrier options}\label{barrier}

We follow the books of Haug \cite{haug_book} and Levy \cite{levy},
and we use the notational structure of Haug.

An up and out call option ceases to exist when
the asset price reaches or goes above the barrier level $B$.
For the up and out call, we follow the formulas
available on p.152--153 of Haug \cite{haug}
(also see (9.77) in Levy \cite{levy}).
These formulas, in turn, originate in the work
of Merton \cite{merton} and Reiner and Rubinstein \cite{rr}.
The up and out call option pays $\max(S-K,0)$ if $S<B$
holds for all times up to $T$, and, otherwise,
it pays a rebate $R$. Then
\begin{align}
  \begin{split}
    C_{K>B} &= F\\
    C_{K<B} &= A - B + C - D + F,
  \end{split}
\end{align}
where
\begin{align}
  \begin{split}
  A &=\phi Se^{(b-r)T}\Phi(\phi x_1)-\phi K e^{-rT}\Phi(\phi x_1-
      \phi\sigma\sqrt{T})\\
    B&=\phi Se^{(b-r)T}\Phi(\phi x_2)-\phi K e^{-rT}\Phi(\phi x_2-
      \phi\sigma\sqrt{T}) \\
    C&=\phi Se^{(b-r)T}\left(\frac{B}{S}\right)^{2(\mu+1)}\Phi(\eta y_1)
      -\phi K e^{-rT}\left(\frac{B}{S}\right)^{2\mu}\Phi(\eta y_1-\eta
      \sigma\sqrt{T})\\
    D&=\phi Se^{(b-r)T}\left(\frac{B}{S}\right)^{2(\mu+1)}\Phi(\eta y_2)
      -\phi K e^{-rT}\left(\frac{B}{S}\right)^{2\mu}\Phi(\eta y_2-\eta
      \sigma\sqrt{T}) \\
   F&= R\left[\left(\frac{B}{S}\right)^{\mu+\lambda}\Phi(\eta z)
        +\left(\frac{B}{S}\right)^{\mu-\lambda}\Phi(\eta z-2\eta\lambda
        \sigma \sqrt{T})\right],
   \end{split}
\end{align}
and
\begin{align}
     x_1&= \frac{\ln(S/K)}{\sigma\sqrt T}+(1+\mu)\sigma\sqrt T &
    x_2&= \frac{\ln(S/B)}{\sigma\sqrt T}+(1+\mu)\sigma\sqrt T\\
  y_1&= \frac{\ln(B^2/(SK))}{\sigma\sqrt T}+(1+\mu)\sigma\sqrt T &
  y_2&= \frac{\ln(B/S)}{\sigma\sqrt T}+(1+\mu)\sigma\sqrt T\\
  z&=\frac{\ln(B/S)}{\sigma\sqrt T}+\lambda\sigma\sqrt T & &\\
  \mu&=\frac{b-\frac{\sigma^2}{2}}{\sigma^2} & \lambda&=
  \sqrt{\mu^2 + \frac{2r}{\sigma^2}}.
\end{align}
\section{Performance}\label{performance}

We compare the Hybrid VA approach,
applying the adjusted spot, strike, and
volatility parameters described in Section \ref{hybrid}
to the barrier option formulas in Section \ref{barrier},
with some work from Dai and Chiu \cite{dai}.
The Hybrid VA and HVA Error columns are new,
and the other columns first appear in the work
of Dai and Chiu.

The details of this test example appear first in their Figure 2
on page 1377, and it provides the default values for
our later comparisons.
There, the risk-free rate is $3\%$, the volatility is $20\%$,
the strike price is $50$, the barrier is $65$, and the time
to maturity is $1$ year. A discrete dividend $1$ is paid
at $0.5$ year. We use Maximum Absolute Error and Root-Mean-Squared Error
as performance indicators.
\begin{table}[ht]
  \centering
  \begin{tabular}{c|c|c|c|c|c|c|c}
    $S(0)$ & MC & Dai-Chiu & Model1 & Hybrid VA & DC Error &
    M1 Error & HVA Error\\\hline
46 & 1.1265 & 1.1260 & 1.1336 & 1.0717 & 0.0005 & 0.0071 & 0.0548\\
48 & 1.3456 & 1.3427 & 1.3641 & 1.2736 & 0.0029 & 0.0184 & 0.0720\\
50 & 1.5054 & 1.5026 & 1.5417 & 1.4219 & 0.0028 & 0.0363 & 0.0835\\
52 & 1.5829 & 1.5796 & 1.6401 & 1.4938 & 0.0033 & 0.0572 & 0.0891\\
54 & 1.5661 & 1.5571 & 1.6422 & 1.4758 & 0.0089 & 0.0762 & 0.0903\\
56 & 1.4389 & 1.4310 & 1.5423 & 1.3649 & 0.0079 & 0.1034 & 0.0740\\
58 & 1.2112 & 1.2093 & 1.3463 & 1.1686 & 0.0019 & 0.1352 & 0.0426\\
60 & 0.9164 & 0.9106 & 1.0700 & 0.9030 & 0.0059 & 0.1536 & 0.0134\\
62 & 0.5667 & 0.5602 & 0.7358 & 0.5898 & 0.0065 & 0.1691 & 0.0231\\
64 & 0.1932 & 0.1868 & 0.3697 & 0.2529 & 0.0065 & 0.1765 & 0.0597\\
   &        &        &        &        & &       &       \\
MAE&        &        &        &        & 0.0089 & 0.1765 & 0.0903\\
RMSE&       &        &        &        & 0.0054 & 0.1109 & 0.0654\\
  \end{tabular}
  \caption{Varying initial stock price for barrier call, single
  discrete dividend}
	\label{tab: S(0)single}
\end{table}

As shown in Table \ref{tab: S(0)single}, as the initial stock price, $S(0)$,
increased, the Model1 formula tended to increase in error as compared with
the HVA model, which maintained some stability in its error magnitude.
The RMSE and MAE of each model's performance indicate this as well.
\begin{table}[ht]
\centering
  \begin{tabular}{c|c|c|c|c|c|c|c}
    $d_1$ & MC & Dai-Chiu & Model1 & Hybrid VA & DC Error
    & M1 Error & HVA Error\\\hline
0.3 & 1.5759 & 1.5730 & 1.5857 & 1.5467 & 0.0029 & 0.0098 & 0.0292\\
0.6 & 1.5438 & 1.5435 & 1.5680 & 1.4928 & 0.0003 & 0.0242 & 0.0510\\
0.9 & 1.5202 & 1.5129 & 1.5486 & 1.4395 & 0.0073 & 0.0283 & 0.0807\\
1.2 & 1.4868 & 1.4815 & 1.5273 & 1.3870 & 0.0053 & 0.0405 & 0.0998\\
1.5 & 1.4478 & 1.4493 & 1.5044 & 1.3353 & 0.0015 & 0.0566 & 0.1125\\
1.8 & 1.4147 & 1.4163 & 1.4798 & 1.2844 & 0.0017 & 0.0652 & 0.1303\\
2.1 & 1.3843 & 1.3828 & 1.4538 & 1.2345 & 0.0015 & 0.0695 & 0.1498\\ 
2.4 & 1.3459 & 1.3488 & 1.4262 & 1.1855 & 0.0030 & 0.0804 & 0.1604\\
    &        &        &        &        & &        &       \\
MAE &        &        &        &        & 0.0073 & 0.0804 & 0.1604\\
RMSE&        &        &        &        & 0.0036 & 0.0523 & 0.1105\\
  \end{tabular}
  \caption{Varying payout amounts for barrier call, single discrete dividend}
	\label{tab: d_1single}
\end{table}

However, with Table \ref{tab: d_1single}, the HVA error in this case
increased while the Model1 error remained relatively flat
as the payout amounts increased.
The MAE and RMSE of each show a drastic difference, with the HVA error
reflecting around double the size in comparison with the Model1 error.
\begin{table}[ht]
\centering
  \begin{tabular}{c|c|c|c|c|c|c|c}
    $\sigma$ & MC & Dai-Chiu & Model1 & Hybrid VA & DC Error &
    M1 Error & HVA Error\\\hline
0.1 & 2.0707 & 2.0756 & 2.0612 & 2.0412 & 0.0049 & 0.0094 & 0.0295\\
0.2 & 1.5054 & 1.5026 & 1.5417 & 1.4219 & 0.0028 & 0.0363 & 0.0835\\
0.3 & 0.7215 & 0.7167 & 0.7534 & 0.6742 & 0.0047 & 0.0320 & 0.0473\\
0.4 & 0.3625 & 0.3611 & 0.3846 & 0.3395 & 0.0014 & 0.0221 & 0.0230\\
0.5 & 0.2035 & 0.1998 & 0.2144 & 0.1881 & 0.0037 & 0.0109 & 0.0154\\
0.6 & 0.1205 & 0.1197 & 0.1292 & 0.1129 & 0.0007 & 0.0087 & 0.0076\\
0.7 & 0.0767 & 0.0764 & 0.0827 & 0.0721 & 0.0003 & 0.0060 & 0.0046\\
0.8 & 0.0526 & 0.0511 & 0.0556 & 0.0483 & 0.0014 & 0.0030 & 0.0042\\
0.9 & 0.0366 & 0.0356 & 0.0388 & 0.0337 & 0.0010 & 0.0021 & 0.0001\\
1.0 & 0.0255 & 0.0255 & 0.0279 & 0.0242 & 0.0000 & 0.0024 & 0.0013\\
    &        &        &        &        & &       &       \\
MAE &        &        &        &        & 0.0049 & 0.0363 & 0.0835\\
RMSE&        &        &        &        & 0.0027 & 0.0178 & 0.0331\\
  \end{tabular}
	\caption{Varying volatility for barrier call, single discrete dividend}
	\label{tab: volsingle}
\end{table}

In Table \ref{tab: volsingle}, both models performed well, with the
Model1 approach doing slightly better.
\begin{table}[ht]
\centering
  \begin{tabular}{c|c|c|c|c|c|c|c}
    $t_1$ & MC & Dai-Chiu & Model1 & Hybrid VA & DC Error &
    M1 Error & HVA Error \\\hline
0.1 & 1.5425 & 1.5378 & 1.5408 & 1.5165 & 0.0047 & 0.0016 & 0.0260\\
0.2 & 1.5347 & 1.5335 & 1.5410 & 1.4926 & 0.0012 & 0.0063 & 0.0401\\
0.3 & 1.5291 & 1.5262 & 1.5412 & 1.4689 & 0.0029 & 0.0121 & 0.0602\\
0.4 & 1.5236 & 1.5160 & 1.5415 & 1.4453 & 0.0076 & 0.0179 & 0.0783\\
0.5 & 1.5054 & 1.5026 & 1.5417 & 1.4219 & 0.0028 & 0.0363 & 0.0835\\
0.6 & 1.4903 & 1.4861 & 1.5419 & 1.3986 & 0.0042 & 0.0516 & 0.0917\\
0.7 & 1.4737 & 1.4658 & 1.5421 & 1.3756 & 0.0079 & 0.0684 & 0.0981\\
0.8 & 1.4391 & 1.4399 & 1.5423 & 1.3527 & 0.0007 & 0.1032 & 0.0864\\
0.9 & 1.4036 & 1.4029 & 1.5425 & 1.3300 & 0.0007 & 0.1389 & 0.0736\\
    &        &        &        &        & &       &       \\
MAE &        &        &        &        & 0.0079 & 0.1389 & 0.0981\\
RMSE&        &        &        &        & 0.0042 & 0.0625 & 0.0745\\
      \end{tabular}
  \caption{Varying payout times for a barrier call,
    single discrete dividend}
			\label{tab: t_1single}
\end{table}

Table \ref{tab: t_1single} shows
that increasing the date of payouts does not yield much better
model performance when evaluating either HVA or Model1.
The RMSE of each are almost identical, and the MAE values
are not much different.
\begin{table}[ht]
\centering
  \begin{tabular}{c|c|c|c|c|c|c|c}
    $S(0)$ & MC & Dai-Chiu & Model1 & Hybrid VA & DC Error &
    M1 Error & HVA Error \\\hline
46 & 0.9156 &  0.9122 & 0.9493 & 0.8126 & 0.0034 & 0.0338 & 0.1030\\
48 & 1.0033 & 1.0028 & 1.0619 & 0.8907 & 0.0005 & 0.0586 & 0.1126\\
50 & 1.0538 & 1.0493 & 1.1322 & 0.9307 & 0.0045 & 0.0783 & 0.1231\\
52 & 1.0484 & 1.0438 & 1.1519 & 0.9276 & 0.0046 & 0.1035 & 0.1208\\
54 & 0.9880 & 0.9843 & 1.1178 & 0.8806 & 0.0037 & 0.1298 & 0.1074\\
56 & 0.8771 & 0.8737 & 1.0316 & 0.7924 & 0.0035 & 0.1545 & 0.0847\\
58 & 0.7241 & 0.7192 & 0.8990 & 0.6692 & 0.0049 & 0.1749 & 0.0549\\
60 & 0.5364 & 0.5315 & 0.7287 & 0.5189 & 0.0049 & 0.1924 & 0.0175\\
62 & 0.3249 & 0.3233 & 0.5317 & 0.3508 & 0.0016 & 0.2068 & 0.0259\\
64 & 0.1104 & 0.1077 & 0.3192 & 0.1744 & 0.0032 & 0.2088 & 0.0640\\
   &        &        &        &        & &        &       \\
MAE&        &        &        &        & 0.0049 & 0.2088 & 0.1231\\
RMSE&       &        &        &        & 0.0038 & 0.1470 & 0.0893\\
      \end{tabular}
  \caption{Varying initial stock price for barrier call,
  two discrete dividends}
			\label{tab: S(0)double}
\end{table}

When studying the difference between the two models, Model1 and HVA,
in the situation of changing $S(0)$ for a barrier call with two dividend
payouts, the differences in the errors become apparent.  Both performed
poorly, but the HVA error was smaller in magnitude.
\begin{table}[ht]
\centering
  \begin{tabular}{c|c|c|c|c|c|c|c}
    $d_1=d_2$ & MC & Dai-Chiu & Model1 & Hybrid VA & DC Error &
    M1 Error & HVA Error\\\hline
0.3 & 1.1305 & 1.1238 & 1.1514 & 1.0841 & 0.0067 & 0.0210 & 0.0464\\
0.6 & 1.0948 & 1.0933 & 1.1462 & 1.0173 & 0.0015 & 0.0514 & 0.0775\\
0.9 & 1.0585 & 1.0606 & 1.1364 & 0.9521 & 0.0021 & 0.0780 & 0.1064\\
1.2 & 1.0279 & 1.0269 & 1.1222 & 0.8885 & 0.0010 & 0.0943 & 0.1394\\
1.5 & 0.9864 & 0.9897 & 1.1038 & 0.8267 & 0.0033 & 0.1174 & 0.1597\\
1.8 & 0.9552 & 0.9535 & 1.0812 & 0.7670 & 0.0018 & 0.1260 & 0.1882\\
2.1 & 0.9147 & 0.9156 & 1.0548 & 0.7095 & 0.0009 & 0.1401 & 0.2052\\
2.4 & 0.8769 & 0.8772 & 1.0247 & 0.6542 & 0.0003 & 0.1478 & 0.2227\\
    &        &        &        &        & &        &       \\
MAE &        &        &        &        & 0.0068 & 0.1478 & 0.2227\\
RMSE&        &        &        &        & 0.0029 & 0.1056 & 0.1547\\
  \end{tabular}
  \caption{Varying payout amounts for barrier call, two discrete
  dividends}
	\label{tab: d_1=d_2}
\end{table}

When changing the payout amounts for two discrete dividend payouts in a
barrier call, again both did poorly, but, in this instance, in contrast with
Table \ref{tab: d_1=d_2} just before, Model1 yielded tighter results.
\section{Further Work}

The tables in Section \ref{performance} show that
the method of Dai and Chiu handily outperforms
the Hybrid VA method. It can also be seen that
the Model 1 approach and the Hybrid VA approach
each have strengths in certain parameter regions,
but neither really compete with the method of Dai and Chiu.

The paper of Buryak and Guo \cite{bg} introduces
the Hybrid VA to price European calls and puts,
and, for that purpose, the method performs reasonably well.
Because their approach doesn't see or make use of barrier information,
it should not be surprising that the modified spot,
strike, and volatility information on their own
do not perform as well for barrier options.
Further, the sharp results of the method of Dai
and Chiu, while derived with solid theoretical
justification, also require non-trivial calculations
to deduce. In particular, their paper \cite{dai}
treats only the single- and double-dividend cases in detail.

It would be interesting to modify or to refine
the Hybrid VA method in a manner that shows greater
sensitivity to the context of barrier options,
say by incorporating the barrier value. Ideally,
an improvement would retain the analytic flavor
of the existing Hybrid VA method of Buryak and Guo
and would achieve results closer to the method
of Dai and Chiu (and to numerical benchmarks
like Monte Carlo methods or Crank-Nicolson schemes).


\end{document}